\documentclass[prl,twocolumn,showpacs]{revtex4}

\usepackage{amsmath,amsfonts,amssymb,amsthm}
\usepackage{graphicx}
\usepackage[active]{srcltx}

\usepackage{color}

\newtheorem{thm}{Theorem}

\newtheorem{assump}{Assumption}

\newcommand{\dist}{{\, \rm dist}}

\newcommand{\C}{\mathbb C}

\newcommand{\di}{\mathrm d}

\theoremstyle{definition}

\begin{document}

\title{Adiabatic quantum
computation:  Enthusiast
 and Sceptic's perspectives}

\author{Zhenwei Cao}
\author{Alexander Elgart}

\affiliation{%
Virginia Tech
}%

\date{\today}

\begin{abstract}
Enthusiast's perspective: We analyze the effectiveness of AQC for a
small rank problem Hamiltonian $H_F$ with the {\it arbitrary}
initial Hamiltonian $H_I$. We prove that for the generic $H_I$ the
running time cannot be smaller than $O(\sqrt N)$, where $N$ is a
dimension of the Hilbert space. We also
 construct an explicit $H_I$ for which the running time is indeed $O(\sqrt
N)$. Our algorithm can be used to solve the unstructured search
problem with the {\it unknown} number of marked items.
\par
Sceptic's perspective: We show that for a {\it robust} device, the
running time for such $H_F$ cannot be much smaller than $O( N/\ln
N)$.
\end{abstract}

\maketitle

{\em Overture}.---Adiabatic quantum computation (AQC)
 ({\it e.g.} \cite{FGGS}) is a
Hamiltonian­-based model of quantum computation. The idea behind AQC
is that finding the ground state of a problem Hamiltonian $H_F$
solves interesting computational problems. In the abstract setting,
let $H_{I}\,,H_{F}$ be a pair of hermitian $N\times N$ matrices,
with $N\gg1$. Consider the interpolating Hamiltonian $H(s)$ of the
form
\begin{equation}\label{eq:H}
H(s):=(1-f(s))H_I+f(s)H_F\,,
\end{equation}
where $f$ is a monotone function on $[0,1]$ satisfying $f(0)=0$,
$f(1)=1$. We will denote by $P_I$ (respectively $P_F$) the spectral
projection on the ground state energy $E_I$ ($E_F$) of the matrix
$H_I$ ($H_F$). We prepare the initial state of the system $\psi(0)$
in the ({\it a-priori} known) ground state $\psi_I\in Range\, P_I$
of the Hamiltonian $H_I$ \cite{range}, and let the system evolve
according to the (scaled) Schr\"odinger equation:
\begin{equation}\label{eq:Sch}
i\dot\psi_\tau(s) \ = \ \tau H(s) \psi_\tau(s)\,,\quad \psi_\tau(0)\
= \ \psi_I\,.
\end{equation}
The adiabatic theorem of quantum mechanics ensures that under
certain conditions the evolution $\psi_\tau(1)$ of the initial state
stays close to the $Range\, P_F$. For AQC to be a potent quantum
algorithm, the running ({\it i.e.} physical) time $\tau$ in
Eq.~\eqref{eq:Sch} must be much smaller than $N$. Although AQC
attracted a considerable interest in physics and computer science
communities, the quantitative characterization of the speed up in
its use remains at large unknown. The core issue here is related to
the extreme sensitivity of the adiabatic behavior to the spectral
structure of the operator $H(s)$. Specifically, the deviations {\it
may} become large when the gap $g(s)$ between the ground state of
$H(s)$ and the rest of its spectrum is small in the vicinity of some
instant $s\in[0,1]$.

The traditional approach to the problem so far was to estimate this
minimal gap \cite{rema}. Putting a few rare exceptions aside ({\it
e.g.} \cite{DMV}), it is usually a very hard task. This explains
why, generally speaking, not much light was shed on the
effectiveness of AQC. Let us note that the estimates of the running
time involving the gap $g(s)$ alone provide only the upper bound on
the optimal running time $\tau$. In reality $\tau$ can be much
smaller.

In this paper, we discuss the reliable upper and lower bounds on
the optimal value of $\tau$, circumventing the estimates on the size
of the gap. Our method is applicable for a particular class of
problem Hamiltonians, satisfying the following hypothesis.
\begin{assump}\label{assump'}
The problem Hamiltonian is of the small rank: $Rank(H_F):=m\ \ll N$.
\end{assump}
Even in this narrower context, there is no unequivocal riposte to
whether AQC is indeed efficient, as we shall see. As often happens
in theoretical deliberations, the answer depends, to some extent, on
the degree of your zeal. To keep the discussion balanced, we present
two different perspectives: The first one is on the optimistic side
while the second  one is rather pessimistic in its nature. To this
end we set a stage for two close acquaintances, Messrs. Enthusiast
and Sceptic, and let the wise Reader judge who of them is closer to
the mark.

Let us note that for AQC to work, it suffices to ensure that
$\psi_\tau(1)$ has a non trivial overlap with the range of $H_F$,
which we will encode in the requirement $\|P_F\psi_\tau(1)\|\ge1/5$,
\cite{remar}.  Another issue that usually arouses certain degree of
confusion, which we  want to avoid, is a normalization of $H(s)$. To
that end, we will use the calibration $\|H_I\|=\|H_F\|=1$. One
should bear this convention in mind when performing comparison with
other results.

The rest of  the paper is organized as follows: We first present the
discussion from Enthusiast and Sceptic's points of view, indicating
briefly the intuition behind the corresponding assertions. We then
give proofs of Theorems \ref{thm:delta} and \ref{thm:at3} (the rest
of the proofs can be found in \cite{CE}). Now we pass the baton to
Mr. Enthusiast.
\par
{\em Enthusiast's perspective}.---To formulate the result,
let me introduce a set of the related parameters. First, I want to
quantify the overlap between the initial state $\psi_I$ and the
problem Hamiltonian. Namely, let $\delta_1 =\| H_F\psi_I\|$, let
$\delta_2 =\| P_F\psi_I\|$, and let $\delta_3 =\| Q_F\psi_I\|$,
where $Q_F$ is a projection onto $Range\, H_F$. Note that for a
generic $H_I$ all $\delta$'s are small, with $\delta_1$ and
$\delta_3$ being $O(\sqrt{m/N})$, while $\delta_2=O(\sqrt{m'/N})$.
Here $m'$ is a dimension of $Range\, P_F$. Second, I want to
distinguish between $E_F$ and the rest of the spectrum of $H_F$,
which I will assume henceforth are separated by the gap $g_F$.
Finally, since I don't want to assume that $H_F$ is sign definite
and given that $\|H_F\|=1$ by convention, the energy $E_F$ will show
up in the estimates.  The prototypical example covered by our
results is the generalized unstructured search (GUS) problem, which
can be cast in the following form: Suppose $H_F$ is diagonal with
the {\it unknown} number of entries equal to $-1$ and the rest of
the entries equal to zero (so that $H_F={\bf 1}-P_F$). Pick
$H_I=-|\psi_I\rangle\langle\psi_I|$ with
$\psi_I=N^{-1/2}(1,\ldots,1)$. Then the corresponding parameters are
$\delta_3 =\delta_2 = \sqrt{m/N}$, $E_F=-1$, and $g_F=1$.

The pair of results below, coupled together, gives fairly tight
lower and respectively upper bounds on the optimal running time in
AQC.
\begin{thm}\label{thm:at1}
Consider the interpolating family Eq.~\eqref{eq:H} with an arbitrary
 $f$. Then the running time $\tau_-$ in Eq.~\eqref{eq:Sch} for
which $\|Q_F\psi_{\tau_-}(1)\|\ge1/5$ satisfies
\begin{equation}\label{eq:min_time}
\tau_-\ \ge \ \frac{1-5\delta_2}{5\delta_1}\,,\quad \mbox{ for
}\quad\delta_2<1/5\,.
\end{equation}
\end{thm}
The quantitative measure of how much $\psi_\tau(s)$ deviates from
$\psi_I$ is encoded in the size of the commutator $[P_I,H(s)]$.
Hence one expects to see the deviation from $\psi_I$ over the time
$\tau$ such that $\tau \cdot\|[P_I,H(s)]\| = O(1)$. Since $P_I$
commutes with $H_I$ while $\|[P_I,H_F]\|\le 2\delta_1$, we get
$\|[P_I,H(s)]\|\le 2\delta_1$ for all $s$ and the bound in
Eq.~\eqref{eq:min_time} follows up to a constant.

Let me note that the similar, albeit less sharp (with the wrong
dependence on $m$) lower bound was recently established in
\cite{IM}.
\begin{thm}\label{thm:at2}
Suppose $\delta_3/g_F=O(1/\ln N)$,
%
%
Then there exists an explicit rank one  $H_I$ and an explicit
function $f$ such that $\|P_F\psi_{\tau_+}(1)\|\ \ge \ 1/5$
for
\begin{equation}\label{eq:min_time'}
\tau_+\ = \
\frac{C(1-E_F)}{|E_F|\,\delta_2}\,,
\end{equation}
for any $C\in[1/3,2/3]$.
\end{thm}
For $N\gg m$ the requirement on $\delta_3/g_F$ is typically
satisfied. Note also that $\tau_-/\tau_+\approx\sqrt{m'/m}$. This is
not particularly surprising, as in Theorem \ref{thm:at1} the aim was
to ensure that $\psi_{\tau_-}(1)$ has an overlap with the range of
$H_F$, whereas in Theorem \ref{thm:at2} we want $\psi_{\tau_+}(1)$
 to overlap with $Range\, P_F$.

The choices in the theorem are: $H_I=- |\psi_I\rangle\langle\psi_I|$
and a (non adiabatic) parametrization $f(s)$ is given by
\[
f(s) \ = \ \begin{cases}
&0\,,\hspace{1.7cm}s=0 \\
 &   \alpha\equiv\frac{1}{1-E_F}\,,\quad s\in(0,1)\\
 &1\,,\hspace{1.7cm} s=1
\end{cases}\,.
\]
That means we move diabatically (instantly) to the given point of
the path, stay there for the time $\tau_+$, and then move quickly
again to the end of the path. Such $f$ is in fact optimal for the
Grover's problem.
\par
The intuition behind this assertion is as follows: With the above
choice for $f(s)$
\[\psi_\tau(1)\ = \ e^{-i\alpha\tau_+\cdot(E_FP_I\,+\,H_F)}\psi_I.\]
Note now that the ground state energy of $E_F P_I$ matches that of
$H_F$ and differs from the energies of its excited states. Let $X$
be a subspace spanned by vectors in the ranges of $P_I$ and $P_F$,
and let $X^\perp$ be its orthogonal complement (so that $X\bigoplus
X^\perp$ is the whole Hilbert space). As usual in adiabatic setting,
the transitions between $X$ and $X^\perp$ are suppressed due to fast
oscillations caused by the energy differential. Therefore the
initial state $\psi_I$  slowly precesses in the $X$ subspace, and by
choosing the right value for $\tau_+$ one can find the evolved state
sufficiently close to $Range\,P_F$. The argument identical to the
one in Theorem \ref{thm:at1} shows that the running time $\tau_+$ is
roughly
\[\frac{1-E_F}{|E_F|}\cdot\frac{1}{\|[P_I,P_F]\|}\,=\,
\frac{1-E_F}{|E_F|}\cdot\frac{1}{2\delta_2}\,.\]
Since the precession is very slow, $\tau_+$ is fairly robust.

This assertion can be seen as an extension of the classical result
of Farhi--Gutmann \cite{FG} on the Grover's search problem. For GUS
the parallel result was established for the quantum circuit model
(QCM) in \cite{BHT}.

\par
Theorem \ref{thm:at2} uses  values of $E_F$ and $\delta_2 =\|
P_F\psi_I\|=O(1/\sqrt N)$ as the input. In many important
applications (such as GUS) the value of $\delta_2$ is unknown. To
this end, we prove the following assertion.
\begin{thm}\label{thm:delta}
Suppose that the value of $E_F$ is known. Then there is a
Hamiltonian -- based algorithm that determines $\delta_2$ with $1/
N^2$ accuracy and requires $O((\ln N)^2)$ of the running time.
\end{thm}
Note that the combined running time in Theorems \ref{thm:at2} and
\ref{thm:delta} remains $O(\sqrt N)$. The algorithm used in the
proof is inspired by the mean ergodic theorem and makes use of the
fact that the survival probability
$c_F(t)=\langle\psi_I|e^{itH_F}\psi_I\rangle$ is directly measurable
in AQC framework. For GUS this problem is known as quantum counting
and was analyzed in QCM framework in \cite{BHT}.
\par
{\em Enthusiast's summary}.---Theorem \ref{thm:at1}
tells us that for a generic $H_I$ the running time cannot be smaller
than $O(\sqrt N)$. Theorems \ref{thm:at2} and \ref{thm:delta}
construct the explicit $H_I$ and the parametrization $f(s)$ so that
$\tau=O(\sqrt N)$. I have assumed that the ground state energy $E_F$
of $H_F$ is known with the $1/N$ accuracy.\par
{\em Sceptic's perspective}.---Let me first point out two
shortcomings
of the method which is usually employed in estimation of the running
time of AQC ({\it e.g.} \cite{DMV} for the Grover's problem and
\cite{RKHLZ}). The technique hinges on a choice of a parametrization
$f$ such that $\dot f(s)$ is small whenever the instantaneous
spectral gap $g(s)$ is small \cite{rem1}. To construct such
$f$, one need to know the values $s_j$ for which
$g(s_j)=O(N^{-1/2})$ with high precision. Such analysis
requires the detailed information about the spectral structure of
$H_F$. The similar issue is present (albeit to a lesser extent) in
the Enthusiast's approach, as one still needs to know $E_F$. Even if
this technical hurdle can be overcome,  the extreme susceptibility
of $\psi_\tau(1)$  to the parametrization $f$ poses a radical
problem in practical implementation.  Indeed, it is presumably
extremely difficult to enforce $\dot f=0$ for a long stretch of the
physical time, as the realistic computing device inevitably
fluctuates. So in the robust setting one can assume that for any
given moment $s_0$ the value $\dot f(s_0)$ is greater than some
small but fixed $\kappa$. We can then as well consider the functions
$f$ in the robust setting that satisfy $\dot f(s)>\kappa$ for {\it
all} values of $s$.

\begin{figure}
  \begin{center}
    \includegraphics[width=7.2cm]{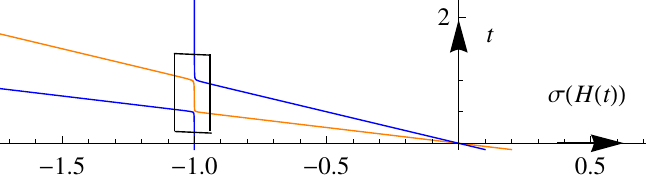}
    \includegraphics[height=3.1cm]{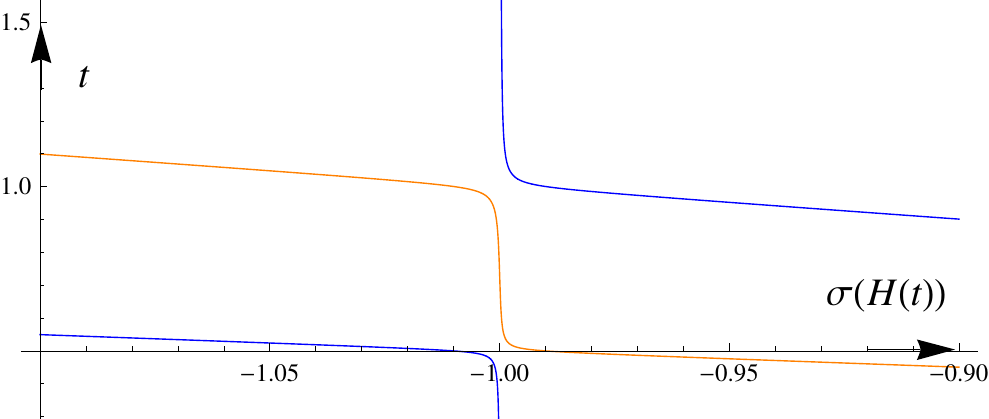}
  \end{center}
  \caption{Illustration for Theorem \ref{thm:at3}, $N=10^4$.
  Top: A pair of relevant eigenvalues of
  $H(t):=H_I+2tH_F$ as a function of $t$, for
$H_F=-\lvert e_1\rangle \langle e_1\rvert - \frac{1}{2}\lvert
e_2\rangle \langle e_2\rvert$ and $H_I=\lvert \phi_1 \rangle \langle
\phi_1 \rvert - \lvert \phi_2 \rangle \langle \phi_2 \rvert$. Here
$e_1=(1,0\ldots,0)$, $e_2=(0,1,0,\ldots,0)$, $\phi_1=1/\sqrt N
(1,\cdots,1)$, and $\phi_2=1/\sqrt N (1,\ldots,1, -1,\ldots,-1)$,
with exactly half of $-1$ so that $\langle \phi_1 \vert \phi_2
\rangle=0$. Bottom: The magnified region of the avoided crossings in
the upper panel.}\label{fig1}
\end{figure}

To understand how the robust system evolves, let me consider the
following semi-empiric argument, substantiated in Theorem
\ref{thm:at3} below. One can show \cite{CE} that for a finite rank
matrix $H_F$ the minimum value $g$ of the gap $g(s)$ between the
ground state energy $E(s)$ of $H(s)$ and the rest of the spectrum of
$H(s)$ is $O(\delta_3)$. Let $g=g(s_0)$, then one can
introduce two different time scales: $\tau_1$ and $\tau_2$. We set
$\tau_1=\Delta^{-1}$, where $\Delta$ is a gap between the two
smallest eigenvalues of $H(s_0)$ and the rest of its spectrum. The
scale $\tau_2$ is associated with a two level system corresponding
to the restriction of the Hilbert space to the spectral subspace of
these two eigenvalues. Typically, $\tau_1\ll\tau_2$, and $\psi_\tau$
stays close to the ground state provided $\tau\gg \tau_2$. If
$\tau_1\ll\tau\ll\tau_2$, then $\psi_\tau$ will still stay close to
the range of the above spectral subspace. However, it  will behave
as if the avoided level crossing is a true level crossing, with
evolution following the first excited state rather than the ground
state (see Figure \ref{fig1}).
To estimate $\tau_2$ let me consider a two level system of the form
\begin{equation}\label{eq:two_level}
K(s)\ = \ \left[\begin{array}{cc}
1-f(s) & \delta f(s) \\
\delta f(s) &f(s)\\
\end{array}\right]\,.
\end{equation}
Assuming that $f$ is differentiable, the Landau--Zener formula shows
that for $K(s)$ the corresponding time scale is $\tau_2\approx \dot
f(s_0)\,\delta^{-2}$. The value of the minimal gap here is equal to
$\delta\sqrt{1+\delta^2}\approx \delta$. Since for $H(s)$ the value
of $g$ is roughly $\delta_3$, we see that  $\psi_\tau$ stays close
to the ground state of $H(s)$ only if
\[\tau\ > \ \dot f(s_0)\cdot(\delta_3)^{-2}\ =\ \dot f(s_0)\cdot O(N)\,.\]
Hence in the robust setting, where $\dot f$ cannot be too small at
any given instant, there is no significant speed up in using AQC.
The result below reaffirms this argument for the case of the initial
Hamiltonian $H_I$ of the small rank. In what follows, $Q_I$ will
denote the projection onto $Range\,H_I$, with $\delta:=\|Q_IQ_F\|$.
\begin{thm}[Robust lower bound on the running time]\label{thm:at3}
Suppose that $f$ in Eq.~\eqref{eq:H} is differentiable and satisfies
$\dot f(s)\ge\kappa>0$ for $s\in[0,1]$. Then, if
$\tau<\tau_r=O\left(\frac{-\kappa}{m^2\delta^2\,\ln\delta}\right)$,
we have
\begin{equation}\label{eq:min_time''}
\left|\langle\psi_I|\psi_\tau\rangle\right|\ >\ \frac{2\sqrt{6}}{5}\,+\,\delta\,.
\end{equation}
Hence the running time $\tau$ for which
$\|Q_F\psi_{\tau}(1)\|\ge1/5$ cannot be smaller than $\tau_r$.
\end{thm}
{\em Sceptic's summary}.---Theorem \ref{thm:at3} tells us that
for a generic $H_I$ of the small rank the robust running time
$\tau_r$ cannot be smaller than $O(N/\ln N)$. Hence AQC is not
really effective for the problem Hamiltonians that satisfy
Assumption \ref{assump'}.
\begin{proof}[Proof of Theorem \ref{thm:delta}]
The starting point is a pair of identities, \cite{GR}:
\begin{eqnarray}\label{eq:iden}
e^{-p}\sum_{k=1}^\infty
\frac{p^k\sin(k\omega)}{k!}&=&e^{p(\cos\omega-1)}\sin(p\sin\omega)\nonumber \\
e^{-p}\sum_{k=1}^\infty
\frac{p^k\cos(k\omega)}{k!}&=&e^{p(\cos\omega-1)}\cos(p\sin\omega)\,.
\end{eqnarray}
In particular, if $1-\cos\omega>\Delta$, each term in
Eq.~\eqref{eq:iden} is bounded by $e^{-p\Delta}$ and therefore is
smaller than $1/ N^2$ provided $p=2\ln N/\Delta$. On the other hand,
the remainders to the partial sums (up to $k=L$) in
Eq.~\eqref{eq:iden} are $O(p^L/L!)$ provided the latter quantity is
small. Combining these observations, we get
\[
e^{-p}\sum_{k=1}^p \frac{p^ke^{ik\omega}}{k!}\,=\,\begin{cases}
&1+O(1/N^2)\,,\quad \omega=0 \\
 &   O(1/N^2)\,,\quad 1-\cos\omega>\Delta
\end{cases}\,,
\]
for  $p=2\ln N/\Delta$ and $\Delta<1$. Hence
\begin{equation}\label{eq:iden'}
e^{-p}\sum_{t=1}^p \frac{p^k}{k!}\,\langle\psi_I|
e^{it(H_F-E_F)}\psi_I\rangle\,=\,(\delta_2)^2+O(1/N^2)\,,
\end{equation}
for $p=2\ln N/(1-\cos g_F)$. The total running time is $\sum_{t=1}^p
t=O((\ln N)^2)$.
\end{proof}
\begin{proof}[Proof of Theorem \ref{thm:at3}]
For a solution $\psi_\tau(s)$ of \eqref{eq:Sch}, let
\begin{equation}\label{eq:dyn_ph}
\phi_\tau(s):= \ e^{ih(s)\tau\,E_I}\psi_\tau(s)\,,\ h(s)=\int_0^s
(1-f(r))\,\di r\,.
\end{equation}
Then one can readily check that $\phi_\tau(s)$ satisfies IVP
\begin{equation}\label{eq:Schd}
i\dot\phi_\tau(s) \ = \ \tau \hat H(s) \phi_\tau(s)\,,\quad
\phi_\tau(0)\ = \ \psi_I\,,
\end{equation}
where $\hat H(s) =  (1-f(s))\,(H_I-E_I)\,+\,f(s)\,H_F$. Clearly
$\left|\langle\psi_I|\psi_\tau\rangle\right|=
\left|\langle\psi_I|\phi_\tau\rangle\right|$. Let
\[
B(s)=\left(f(s)H_F-((1-f(s))E_I+\epsilon i\right))^{-1}\,,
\]
and let $\phi(s) = \psi_I-f(s)H_FB(s)\, \psi_I$, where $\epsilon$ is
a small parameter to be chosen later. Omitting the $s$ dependence,
we have
\begin{equation}\label{eq:hatHp}
\hat H\phi= -f(1-f)H_IH_F B \,\psi_I +i\epsilon fH_FB \,\psi_I\,.
\end{equation}
That means that away from the $m$ values of $s$ for which $B(s)$ has
zero eigenvalue, $\|\hat H\phi\|$ is very small, since
$\|H_IH_F\psi_I\|\le\delta^2$.  Note now that
\begin{equation}\label{eq:ftc}
\langle\phi(1)|\phi_\tau(1)\rangle \ = \
\langle\phi(0)|\phi_\tau(0)\rangle\,+\,\int_0^1\frac{d}{ds}
\langle\phi(s)|\phi_\tau(s)\rangle\,ds\,.
\end{equation}
But $\langle\phi(0)|\phi_\tau(0)\rangle=1$ and
\begin{eqnarray*}
\left|\langle\phi(1)|\phi_\tau(1)\rangle\right|& =&
\left|\langle\psi_I|\phi_\tau(1)\rangle \,-\,
\langle\psi_I|\frac{H_F}{H_F-\epsilon\,i}| \phi_\tau(1)\rangle\right|\\
&&\hspace{-2cm}\le\
\left|\langle\psi_I|\phi_\tau(1)\rangle\right|\,+\,\|Q_F\psi_I\|\
\le \ \left|\langle\psi_I|\phi_\tau(1)\rangle\right|\,+\,\delta\,.
\end{eqnarray*}
Substitution into Eq.~\eqref{eq:ftc} gives
\[1\,-\,\left|\langle\psi_I|\phi_\tau(1)\rangle\right|\ \le \
\left|\int_0^1\frac{d}{ds}
\langle\phi(s)|\phi_\tau(s)\rangle\,ds\right| \,+\,\delta\,.\]
Hence Eq.~\eqref{eq:min_time''} will follow if
\begin{equation}\label{eq:nbn}
\left|\int_0^1\frac{d}{ds}
\langle\phi(s)|\phi_\tau(s)\rangle\,ds\right| \ <\
1\ -\ \frac{2\sqrt{6}}{5}\,-\,2\delta\,.
\end{equation}
We have
\[\frac{d}{ds}
\langle\phi(s)|\phi_\tau(s)\rangle \ = \
\langle\dot\phi(s)|\phi_\tau(s)\rangle \,-\,i\tau\langle\phi(s)|\hat
H(s)|\phi_\tau(s)\rangle\,.\]
We bound the first term on the right hand side by $\|\dot\phi\|$ and
the second one by $\tau\|\hat H\phi_\tau\|$. A straightforward
computation (using Eq.~\eqref{eq:hatHp} for the second term) shows
that
\[\|\dot\phi\| \ \le\ \frac{\dot f\delta}{\Delta_\epsilon}\,+\,
\frac{2\dot f\delta}{\left(\Delta_\epsilon\right)^2}\,;\ \|\hat
H\phi_\tau\| \ \le\
\frac{\delta^2+\epsilon\delta}{\Delta_\epsilon}\,,
\]
where
$\Delta_\epsilon(s):=\dist(f(s)\sigma(H_F)\,,\,(1-f(s))E_I+\epsilon\,i)$.
Here $\dist(S,z)$ is an Euclidean distance from the set $S$  to the
point $z$ in $\C$, and $\sigma(H)$ stands for the spectrum of $H$.

As a result, we obtain a bound
\[\left|\dot{\overbrace{
\langle\phi|\phi_\tau\rangle}}\right| \ \le \ (\dot
f+\tau\delta+\tau\epsilon)\frac{\delta}{\Delta_\epsilon}\,+\,
2\frac{\dot f\delta}{\left(\Delta_\epsilon\right)^2}\,.\]
Integrating both sides over $s$ and using the bounds
\begin{eqnarray*}
\int_0^1\frac{\dot f ds}{\Delta_\epsilon(s)}& \le &
-2m\ln\epsilon\,;\quad \int_0^1\frac{\dot f
ds}{\left(\Delta_\epsilon(s)\right)^2}\ \le \
\frac{2m}{\epsilon}\,;\\ \int_0^1\frac{ds}{\Delta_\epsilon(s)}& \le
& -2m\frac{\ln\epsilon}{\kappa}\,,
\end{eqnarray*}
we can estimate
\begin{multline*}
\left|\int_0^1\frac{d}{ds}
\langle\phi(s)|\phi_\tau(s)\rangle\,ds\right| \\ \le\
2m\delta\left(-\ln\epsilon\left(1+\frac{\tau\delta}{\kappa}+
\frac{\tau\epsilon}{\kappa}\right)+\frac{2}{\epsilon}\right)\,.
\end{multline*}
Hence the required bound in Eq.~\eqref{eq:nbn} follows with the
choice $\epsilon=10^{-3}m\delta$, provided $\tau\le
-\frac{C\kappa}{\epsilon^2\ln\epsilon}$ where $C$ is a constant.
\end{proof}
Partially supported by the NSF Grant DMS--0907165.


\begin{thebibliography}{99}
\bibitem{FGGS}
E.~Farhi {\it et al.}, Science {\bf 292},  472  (2001).

\bibitem{range} The range of the operator $A$ on a Hilbert space
$X$  is a collection of all vectors $y$ such that $y=Ax$ for some
$x\in X$. The rank of $A$ is a dimension of $Range \, A$ and
coincides with the number of non zero eigenvalues for hermitian $A$.
For example, for $A=|\psi\rangle\langle\psi|$ the range consists of
vectors proportional to $\psi$ and $Rank \,A=1$.

\bibitem{rema}
We will relate our results with the prior gap-free bounds on $\tau$
(namely \cite{IM,FG}) in the next section.

\bibitem{DMV}
W.~van Dam {\it et al.}, 42nd IEEE Symposium on Foundations of
Computer Science 279 (2001).

\bibitem{remar}
In fact, it suffices to have an overlap of order $1/p(\ln N)$ ,
where $p(x)$ is a polynomial in $x$.

\bibitem{CE} Z.~Cao and A.~Elgart,  arXiv:1004.4911v1.

\bibitem{IM}
L.~M. Inannou and M.~Mosca, Int. J. Quant. Inform.  {\bf 6}, 419
(2008).

\bibitem{FG}
E.~Farhi and S.~Gutmann, Phys. Rev. A {\bf 57}, 2403 (1998)

\bibitem{BHT}
G.~Brassard {\it et al.}, Lecture Notes in Comput. Sci. {\bf 1444},
820 (Springer-Verlag, New York/Berlin (1998)).

\bibitem{RKHLZ}
A.~T. Rezakhani {\it et al.}, Phys. Rev. Lett. {\bf 103}, 080502
(2009).

\bibitem{rem1}
In fact, the Enthusiast's approach brings this strategy to its
extreme by choosing $\dot f(s)=0$ for {\it all} $s\in(0,1)$. One can
show \cite{CE} that that for the sign definite $H_F$ the number of
avoided crossings is exactly equal to $m$.

\bibitem{GR}
I.~S. Gradshteyn  and I.~M. Ryzhik, formulae (1.449), {\em Table of
integrals, series, and products} (Elsevier/Academic Press,
Amsterdam, 2007).






\end{thebibliography}
\end{document}